\PassOptionsToPackage{sort&compress}{natbib}
\documentclass[aps,prd,twocolumn,superscriptaddress,nofootinbib,floatfix]{revtex4-2}
\usepackage[T1]{fontenc}
\usepackage{mathptmx} 
\usepackage{amsmath,bm,amssymb}
\usepackage{graphicx}
\usepackage{booktabs}
\usepackage{orcidlink}
\usepackage{xcolor}

\usepackage{hyperref}
\hypersetup{colorlinks=true,linkcolor=blue,citecolor=magenta,urlcolor=blue}
\def\doi#1{\href{https://doi.org/#1}{#1}}
\newcommand{\Msun}{M_\odot}
\newcommand{\lam}{\lambda}
\newcommand{\NS}{{\rm NS}}
\newcommand{\BH}{{\rm BH}}
\newcommand{\GB}{Gauss--Bonnet}
\newcommand{\GW}{Gibbons--Werner}

\newcommand{\ISCO}{{\rm ISCO}}
\newcommand{\EoS}{equation of state}
\newcommand{\upi}{\pi}

\newif\ifreview \reviewtrue
\ifreview  \else  \fi

\begin{document}
\title{Scalar charge of neutron stars and accreting black holes in Starobinsky gravity: quasi-universal relations, the trace anomaly and gravitational-wave sensitivity}
\author{Yashmitha Kumaran\,\orcidlink{0009-0006-5655-6931}}
\email{yashmitha@astro.unam.mx}
\affiliation{Instituto de Astronom\'ia, Universidad Nacional Aut\'onoma de M\'exico, AP 70-264, Ciudad de M\'exico 04510, M\'exico}
\author{Il\'idio Lopes\,\orcidlink{0000-0002-5011-9195}}
\email{ilidio.lopes@tecnico.ulisboa.pt}
\affiliation{Centro de Astrof\'isica e Gravita\c{c}\~ao (CENTRA), Departamento de F\'isica, Instituto Superior T\'ecnico (IST), Universidade de Lisboa (UL), Av.~Rovisco Pais~1, 1049-001 Lisboa, Portugal}

\begin{abstract}
The Starobinsky scalaron is sourced by the trace of the stress--energy, so its exterior charge is set by the matter around an object and not by the object's vacuum geometry. We derive a master formula for that charge for any static, spherically symmetric source, evaluate it first for neutron stars, where the interior is known, and then for accreting black holes, where it is not. For a star the pressure subtracts: on the interior metric $q_{\rm curved}=1-6P_T/M=1-2s$, with $s$ the Eardley sensitivity, and the charge is a quasi-universal function of the compactness. Six realistic equations of state confirm the relation but lie above the coefficient calibrated on polytropes; the refit $q_{\rm curved}\simeq(1-\mathcal{C})(1-0.205\,\mathcal{C}-4.61\,\mathcal{C}^2)$ reproduces their mean to $0.007$. The trace $\rho-3p$ changes sign in the cores of massive stars, and the charge also tracks the tidal deformability and the moment of inertia to a few parts in $10^{3}$ and $10^{2}$. Solving the Einstein-frame structure equations self-consistently moves a star along this relation and leaves its charge at fixed compactness within $4$ per cent of the test-field value. We compare the charge with NICER radii and the GW170817 tidal deformability, and use the published massless dipole constraint from GW230529 to calibrate a Fisher estimate of the sensitivity of a similar binary to a finite scalaron range, projected to the Einstein Telescope and Cosmic Explorer; the range $\lambda$ is a free parameter, and the laboratory bound on pure $R+\alpha R^2$ is confronted explicitly. A vacuum black hole has $q=0$; a hole with a disc has $q=\sinh(m_\phi r_d)/(m_\phi r_d)$, controlled by the disc's outer edge, which makes it the counterpart in which the same integral is dominated by matter far from the centre; light is blind to the charge at leading order, and the correction beyond it is negligible for any self-consistent disc.
\end{abstract}
\maketitle

\section{Introduction}
\label{sec:intro}

Neutron stars are the natural laboratory for a scalar degree of freedom sourced by matter. Their interiors are fixed by nuclear physics to within the uncertainty of the equation of state; their masses and radii are now measured by NICER~\cite{Riley2019,Miller_2019,Choudhury:2024xbk,Salmi:2024aum}; and their tidal deformabilities, together with the dipole radiation of binaries that contain them, are measured by gravitational-wave detectors~\cite{LIGOScientific:2017vwq,LIGOScientific:2018cki,LIGOScientific:2024elc}. A scalar coupled to the trace of the stress--energy turns each of these measurements into a test, provided that the charge the star carries can be computed.

That charge is the question we address here. In quadratic $f(R)$ gravity, the Starobinsky model~\citep{Starobinsky1980,Sotiriou2010,DeFelice2010,NojiriOdintsovOikonomou2017}, the scalaron is massive, so its influence is screened beyond a range $\lam=1/m_\phi$, and we have shown elsewhere~\citep{KumaranLopes2026} that this screening severs the exterior of a vacuum black hole from that of a neutron star of the same mass: the black hole carries no scalar charge and bends light exactly as general relativity requires, whilst the star acquires a charge weighted by the pressure that supports it. That treatment left the amplitude of the charge undetermined, parametrising it by a number $q$ whose value, and whose sensitivity to the \EoS, an exterior-only calculation cannot fix. The present article fixes $q$, first for the neutron star, where it can be confronted with data, and then for an accreting black hole, and shows that the same integral delivers it.

The black hole is the instructive counterpart. The no-hair theorems forbid a static, asymptotically flat black hole in vacuum from carrying an independent scalar charge, in scalar-tensor gravity and in $f(R)$ gravity alike~\citep{Hawking1972,SotiriouFaraoni2012,Canate2018,SultanaKazanas2018}. They say nothing, however, about a horizon that sits inside matter. Real black holes are not in vacuum: they are surrounded by accretion flows, and the trace of the stress-energy of that flow is precisely what sources the scalar degree of freedom of a higher-curvature theory. Jacobson~\cite{Jacobson1999} showed that a black hole immersed in an evolving scalar environment acquires a profile fixed entirely by its surroundings, a conclusion since confirmed by numerical evolutions of black holes in scalar-tensor theories with matter~\citep{HorbatschBurgess2012,Berti2013,Cardoso2013}; the same logic applies to a static trace source outside the horizon. We call the resulting field \emph{environmentally sourced} rather than hair for exactly this reason: it belongs to the environment and not to the black hole, so that the theorem is untouched.

Central to the applicability of the whole construction is a single observation. For any static, spherically symmetric source the exterior scalaron is a Yukawa profile whose amplitude is the trace of the stress-energy weighted by $\sinh(m_\phi r)/(m_\phi r)$, a kernel that equals unity at the centre and grows exponentially with radius. Two consequences follow immediately, and they pull in opposite directions. Pressure enters the trace with the opposite sign to density and therefore suppresses the charge of a star; distance from the centre enters through the kernel and therefore enhances the charge of a disc. A neutron star and an accretion disc of the same mass consequently sit on opposite sides of the point-mass benchmark, and it is the environment that decides which.

The closest antecedent of this idea is that of Psaltis et al.~\cite{Psaltis:2007cw}, who showed that although the Kerr metric solves $f(R)$ gravity in vacuum, matter around a black hole can make the solution depart from Kerr; and it sits within a broader literature on environmental effects on black holes and their gravitational waves~\cite{Barausse:2014tra,Cole:2022yzw}. What we add is a massive scalaron, a horizon-regular boundary condition derived from the field equation, and the reduction of the problem to a single number $q$.

The article leads with the neutron star, where the interior is under control and where data exist. Section~\ref{sec:formalism} derives the master formula, fixes the boundary condition at a horizon, and states the scope of the work, including the laboratory bound on the scalaron range (Sec.~\ref{sec:scope}). Section~\ref{sec:ns} evaluates the charge for a star, first for polytropes, then on the interior metric, then for six realistic equations of state (Sec.~\ref{sec:eos}), where the refit of the charge--compactness relation is given; Secs.~\ref{sec:anomaly} and~\ref{sec:universal} give the trace-anomaly structure of the charge and its relation to $\Lambda$ and $\bar I$, and Sec.~\ref{sec:scf} solves the structure equations with the scalaron's back-reaction included. Section~\ref{sec:data} confronts the results with NICER, GW170817 and a GW230529-like dipole sensitivity estimate, and collects the constraints in the $(\lambda,q)$ plane. Section~\ref{sec:bh} treats the accreting black hole as the counterpart in which the same integral is dominated by matter far from the centre: the disc model and its validity, the lensing it implies, its amplitude and the bearing of horizon-scale imaging. The discussion and conclusion close the article.

\section{The environmentally sourced scalaron}
\label{sec:formalism}

For $f(R)=R+\alpha R^2$ the scalar degree of freedom is made explicit by the Legendre map $\phi\equiv f'(R)-1=2\alpha R$ of Whitt~\cite{Whitt1984}, and it obeys the wave equation of a massive field exactly. The trace of the field equations reads $f'(R)R-2f(R)+3\Box f'(R)=8\upi T$, and for this $f$ the algebraic part collapses, $f'R-2f=-R$ identically, so that with $1/2\alpha=3m_\phi^2$ the whole trace becomes the wave equation of $\phi$ sourced by the trace $T$ of the matter stress-energy, 
\begin{equation}
\Box \phi - m_\phi^2 \phi = +\frac{8\upi}{3}\,T , \qquad m_\phi^2 = \frac{1}{6\alpha} ,
\label{eq:field}
\end{equation}
where $\alpha$ is the coupling of the quadratic term and $\lam\equiv1/m_\phi=\sqrt{6\alpha}$ is the scalaron range. Nothing has been dropped in reaching Eq.~(\ref{eq:field}); what is approximated in what follows is the background upon which $\Box$ is built, the field being treated as a test field that does not itself deform the metric. Since $f'(R)=1+\phi$, the amplitude of $\phi$ is also the fractional departure of the effective gravitational coupling from unity, so that $|\phi|$ measures how good that treatment is. For the matter considered here, $T=-\rho+3p$, so that a positive density sources a positive $\phi$ and the pressure opposes it. On any static, spherically symmetric background~\citep{MTW1973},
\begin{equation}
ds^2=-e^{2\Phi(r)}dt^2+e^{2\Lambda(r)}dr^2+r^2d\Omega^2 , \qquad e^{-2\Lambda}=1-\frac{2m(r)}{r} ,
\label{eq:metric}
\end{equation}
Eq.~(\ref{eq:field}) becomes the radial equation
\begin{equation}
\frac{d}{dr}\!\left[e^{\Phi-\Lambda}r^2\frac{d\phi}{dr}\right] - m_\phi^2\,e^{\Phi+\Lambda}r^2\phi = \frac{8\upi}{3}\,e^{\Phi+\Lambda}r^2\,T .
\label{eq:curvedradial}
\end{equation}
Both combinations in Eq.~(\ref{eq:curvedradial}) descend from the determinant, $\sqrt{-g}=e^{\Phi+\Lambda}r^2\sin\theta$ and $\sqrt{-g}\,g^{rr}=e^{\Phi-\Lambda}r^2$, the square root halving the exponents of Eq.~(\ref{eq:metric}). This is the form in which a scalar sourced by a relativistic star has long been treated in scalar-tensor gravity~\citep{DamourEspositoFarese1993}. Two cases are needed. Outside a vacuum horizon $m(r)=M$ and $e^{\Phi}=e^{-\Lambda}=\sqrt{f}$ with $f(r)=1-2M/r$, so that Eq.~(\ref{eq:curvedradial}) reduces to
\begin{equation}
\Box\phi = \frac{1}{r^2}\frac{d}{dr}\!\left[r^2 f(r)\,\frac{d\phi}{dr}\right] = f\,\phi'' + \left(\frac{2}{r}f + \frac{2M}{r^2}\right)\phi' .
\label{eq:box}
\end{equation}
Inside a star $m(r)$ is instead the enclosed mass and $\Phi$ is fixed by the structure equations. The coefficient of $\phi''$ in Eq.~(\ref{eq:box}) vanishes at the horizon, so that $r=2M$ is a regular singular point of the ordinary differential equation and not a physical singularity of $\phi$. The correct boundary condition there is regularity~\citep{Bekenstein1995,HerdeiroRadu2015}, not the flat-space Neumann stand-in $d\phi/dr=0$ that a naive treatment would impose.

\subsection{The sinh-weighted scalar charge}
\label{sec:sinh}

Before turning to the curved background it is worth extracting the mechanism in its simplest setting, since it is what the two environments share. In flat space Eq.~(\ref{eq:field}) is a Helmholtz equation with the standard Yukawa kernel, and integrating it over a static, spherically symmetric source of density $\rho$ and pressure $p$ gives, at any radius outside the matter,
\begin{align}
\phi_{\rm ext}(r) &= \frac{2Q_\phi}{r}\,e^{-m_\phi r} , \notag \\[2pt]
Q_\phi &= \frac{4\upi}{3}\int_0^{R_{\rm src}}(\rho-3p)\, \frac{\sinh(m_\phi r)}{m_\phi r}\,r^2\,dr .
\label{eq:charge}
\end{align}
Equation~(\ref{eq:charge}) exhibits the $\sinh$-weighted trace sourcing in explicit form: the entire mechanism resides in the kernel $\sinh(m_\phi r)/(m_\phi r)$, which weights the trace of the stress-energy at every radius. It is convenient to normalise it by the source mass $M_{\rm src}=4\upi\int\rho\,r^2dr$ and define the dimensionless charge 
\begin{equation}
q \equiv \frac{3Q_\phi}{M_{\rm src}} = \frac{\displaystyle\int(\rho-3p)\,\frac{\sinh(m_\phi r)}{m_\phi r}\,r^2\,dr} {\displaystyle\int\rho\,r^2\,dr} ,
\label{eq:qdef}
\end{equation}
which is unity for a pressureless point mass, since the kernel tends to one as $r\to0$, and which is the parametrisation $q=3Q_\phi/M\in[0,1]$ used for the isolated case by Kumaran \& Lopes~\cite{KumaranLopes2026}. The whole of the present article can be read as the evaluation of Eq.~(\ref{eq:qdef}) in three environments. For a vacuum black hole there is no matter and $q=0$ identically. For a thin shell or ring of matter at radius $r_d$, the kernel is constant across the source and
\begin{equation}
q_{\rm shell} = \frac{\sinh(m_\phi r_d)}{m_\phi r_d} \;\ge\; 1 ,
\label{eq:qshell}
\end{equation}
growing exponentially once the matter lies beyond a scalaron range. For a star, the pressure term subtracts and the two effects compete. Two remarks are worth making at once. First, Eq.~(\ref{eq:qshell}) is the precise sense in which the screening length felt at a radius $r$ is $r-r_d$ and not just $r$: the factor $\sinh(m_\phi r_d)\,e^{-m_\phi r}\to\tfrac12 e^{-m_\phi(r-r_d)}$ for $m_\phi r_d\gg1$, so the exponential is measured from the matter and not from the centre, which is only to be expected of a screened field.  Second, because the kernel grows exponentially, the integral in Eq.~(\ref{eq:charge}) is dominated by the outermost matter whenever the source extends beyond a scalaron range; the outer truncation radius of a disc is therefore a physical parameter of the problem and not a numerical convenience.

\subsection{The kernel on a curved background}
\label{sec:kernel}

The $\sinh$ weighting of Eq.~(\ref{eq:charge}) is not an ansatz, and it is worth saying where it comes from, because the same statement tells us what replaces it when the background is not flat. Equation~(\ref{eq:curvedradial}) is linear, so a source with a decaying exterior is integrated by the Green's function built from the two homogeneous solutions: $u$, regular at the centre and normalised to $u(0)=1$, and $v$, decaying at large radius. The exterior amplitude is then
\begin{align}
Q_\phi &= \frac{4\upi}{3}\,\frac{1}{|W|}\int (\rho-3p)\,u(r)\,e^{\Phi+\Lambda}r^2\,dr , \notag \\[2pt]
W &= e^{\Phi-\Lambda}r^2\left(uv'-u'v\right) ,
\label{eq:master}
\end{align}
the Wronskian $W$ being constant by construction. The kernel that weights the trace is therefore the interior regular mode, and the measure $e^{\Phi+\Lambda}r^2dr$, the proper volume weighted by the lapse, replaces the coordinate one. In flat space the two modes are $\sinh(m_\phi r)/(m_\phi r)$ and $e^{-m_\phi r}/r$, the Wronskian is $-1$, and Eq.~(\ref{eq:master}) returns Eq.~(\ref{eq:charge}) exactly; both are the monopole of the Yukawa multipole expansion of the Green's function~\citep{Olver2010}. Equation~(\ref{eq:master}) therefore holds on any static, spherically symmetric background, and reduces to the flat-space expression whenever the source occupies a region in which the metric is close to flat. Equation~(\ref{eq:qdef}) is untouched by any of this, being a statement about the exterior amplitude and not about how one computes it.

\subsection{Horizon-regular boundary condition}
\label{sec:robin}

Writing $x=r-2M$ and expanding Eq.~(\ref{eq:box}) near $x=0$ gives, to leading order,
\begin{equation}
x\,\phi'' + \phi' - 2M m_\phi^2 \phi = 0 ,
\label{eq:frob}
\end{equation}
a regular singular equation with a double indicial root $s=0$~\citep{BenderOrszag1999}: one solution is analytic in $x$, the other carries a $\ln x$ divergence. We fix the regular branch by solving the Frobenius series order by order, which resums to a modified Bessel function, $\phi\propto I_0\!\left(2\sqrt{2Mm_\phi^2x}\right)$ (the recurrence being $n^2a_n=2Mm_\phi^2a_{n-1}$), and impose the Robin condition
\begin{equation}
\left.\frac{d\phi/dr}{\phi}\right|_{r=r_{\rm in}}
= \mathcal{R}(M,m_\phi,x_{\rm in})
\label{eq:robin}
\end{equation}
at a small offset $x_{\rm in}$ from the horizon, $r_{\rm in}=2M+x_{\rm in}$, in place of the ad hoc Neumann condition. For $M=1$, $m_\phi=0.3$ and $x_{\rm in}=0.05M$ we find $\mathcal{R}\approx0.179$ from $\mathcal{R}=(2Mm_\phi^2/x_{\rm in})^{1/2}\,I_1(z)/I_0(z)$ with $z=2(2Mm_\phi^2x_{\rm in})^{1/2}$, close to its limiting value $2Mm_\phi^2=0.18$ at the horizon itself, and we have confirmed that the solution is stable to better than one per cent under two orders of magnitude of variation in $x_{\rm in}$; the fix does not simply relocate the arbitrariness it removes.

\subsection{Scope: the laboratory bound and what is claimed}
\label{sec:scope}

In pure $R+\alpha R^2$ the scalaron mass is a constant, $m_\phi^2=1/6\alpha$, so there is no chameleon screening, and the Yukawa coupling to matter is fixed at $1/3$. Torsion-balance experiments~\cite{Kapner:2006si,Lee:2020zjt} then bound $|\alpha|\lesssim2\times10^{-9}\,{\rm m}^2$~\cite{Berry:2011pb,Naf:2010zy}, that is $\lam\lesssim10^{-4}$~m. The ranges $\lam\sim30$--$300$~km on which the figures of this article are built lie some nine orders of magnitude beyond that, and the gravitational-wave sensitivity estimate of Sec.~\ref{sec:data_gw}, $\lam\lesssim (3.5\text{--}4)\times10^{2}$~km for the stellar charges $q=0.66$ and $0.49$, is correspondingly weak in comparison. We therefore state what is, and is not, claimed.

We treat $\lam$ as a free parameter of a massive scalar coupled to the trace of the stress--energy with the strength fixed by $f(R)$, for which $R+\alpha R^2$ is the minimal realisation. Everything derived below holds for that field equation, Eq.~(\ref{eq:field}), at any $\lam$ and says what an object in that theory carries; the astrophysical channels are then independent strong-field tests of the same field equation, probing curvatures and compactnesses that no laboratory experiment reaches. Whether a given theory is compatible with the laboratory bound at the same time is a separate question about its UV completion, and we do not claim that pure $R+\alpha R^2$ is. The natural alternative is a wider $f(R)$ family with a density-dependent scalaron mass~\cite{Hu:2007nk,Khoury:2003aq,Brax:2008hh}, in which the range in a laboratory vacuum differs from that inside or around a compact object. We do not adopt it, because in that case the interior of a neutron star is in general screened by a thin shell, the scalaron mass is no longer constant in Eq.~(\ref{eq:curvedradial}), and the master formula and the relations of Sec.~\ref{sec:ns} would have to be rederived; Sec.~\ref{sec:data_eht} notes that low-accretion-rate black holes are in the unscreened regime assumed here~\citep{Frolov:2017asg}. Throughout, the constraints of Sec.~\ref{sec:data} are to be read as constraints on the unscreened scalaron range in that sense.

\section{The neutron star}
\label{sec:ns}

The neutron-star channel differs from the accreting black hole in what sets the charge, and in nothing else: Eq.~(\ref{eq:qdef}) applies unchanged. Neither of the two approximations weighed for the disc in Sec.~\ref{sec:validity} arises for a static, non-rotating star, since hydrostatic equilibrium is exactly stationary and the monopole reduction is exact; nor is a horizon-regular condition of the kind of Sec.~\ref{sec:robin} needed, the interior being regular at the origin by construction. What remains is the interior profile itself, which we now compute.

\subsection{Stellar models and the scalar charge}

We integrate the Tolman--Oppenheimer--Volkoff equations~\citep{Tolman1939,OppenheimerVolkoff1939} for a polytropic \EoS\ $p=K\rho_{\rm b}^\Gamma$ with $\rho=\rho_{\rm b}+p/(\Gamma-1)$, $\rho_{\rm b}$ being the rest-mass density and $\rho$ the total energy density, and evaluate Eq.~(\ref{eq:qdef}) on the resulting profiles. Our fiducial model, $\Gamma=2$ with $K=187\,{\rm km}^2$, has $M=1.4\,\Msun$ and $R_\star=12.4$~km, compactness $\mathcal{C}\equiv M/R_\star=0.167$, which is comfortably inside the range inferred from NICER timing~\citep{Riley2019,Raaijmakers2021}. Writing $q_\infty$ for its charge in the limit of long scalaron range, $\lam\gg R_\star$,
\begin{equation}
q_\infty = 1 - \frac{3}{M}\!\int p\,dV = 1-0.219 = 0.781 ,
\label{eq:qinf}
\end{equation}
the first equality holding because the kernel of Eq.~(\ref{eq:qdef}) reduces to unity in that limit, and the numbers being those of the fiducial star. A neutron star therefore carries about four-fifths of the scalar charge that a point mass of the same mass would carry, the deficit being three times its volume-integrated pressure. This is the number that an exterior-only treatment cannot supply.

\begin{figure*}
\centering
\includegraphics[width=\textwidth]{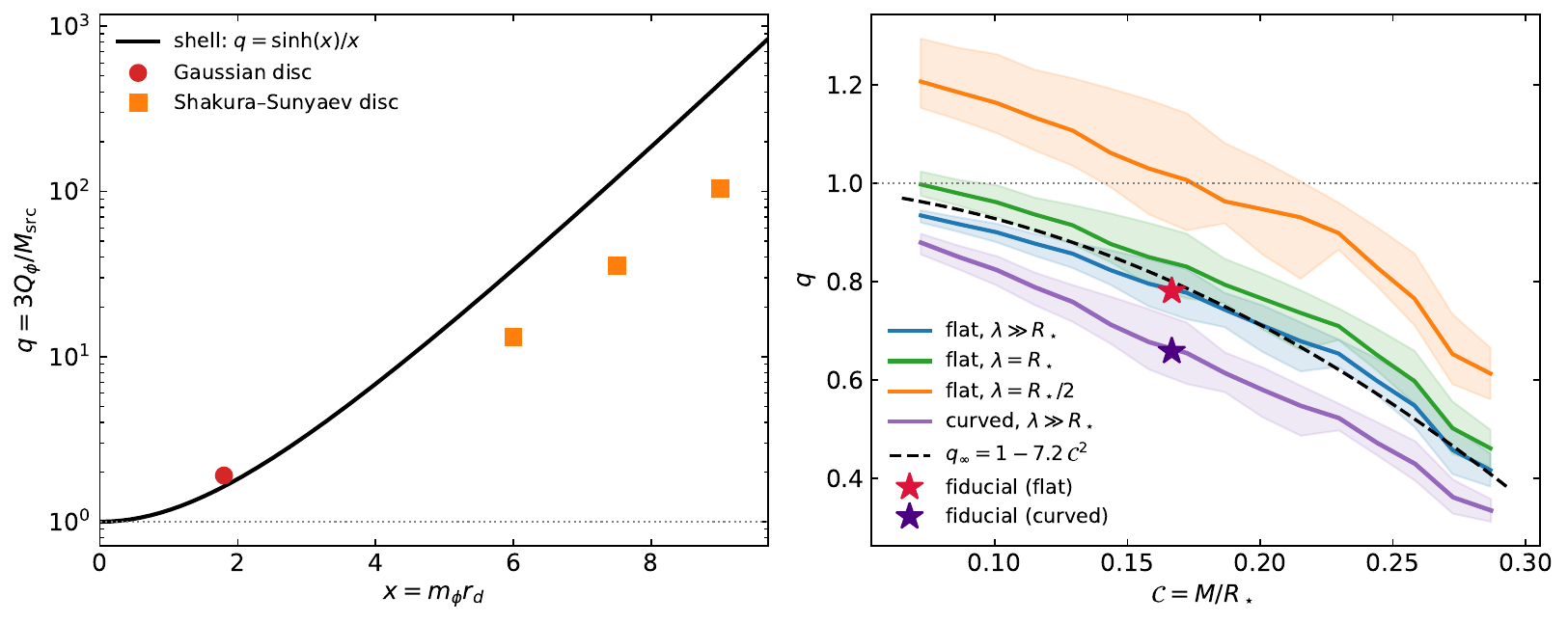}
\caption{The dimensionless scalar charge $q=3Q_\phi/M_{\rm src}$ of Eq.~(\ref{eq:qdef}) in the two environments. (a) Accreting black hole: the exact shell result of Eq.~(\ref{eq:qshell}), $q=\sinh(x)/x$ with $x=m_\phi r_d$, against the Gaussian disc of Sec.~\ref{sec:curvedflat} ($x=1.8$, $q=1.90$) and the Shakura--Sunyaev disc of Sec.~\ref{sec:disk} truncated at $r_{\rm out}=20M$, $25M$, and $30M$ ($q=13.1$, $35.4$, and $104$); the widening gap between the shell curve and the truncated-disc points is the point of the panel. (b) Neutron star: the $150$-model polytrope family of Sec.~\ref{sec:ns}, against compactness $\mathcal{C}=M/R_\star$. Lines are the median charge and shaded bands the full spread across the family, for the flat-space propagator at $\lam\gg R_\star$, $\lam=R_\star$, and $\lam=R_\star/2$, and for the curved-background propagator of Sec.~\ref{sec:nscurved} at $\lam\gg R_\star$; dashed line is the fit of Eq.~(\ref{eq:qfit}), $q_\infty=1-7.2\,\mathcal{C}^2$; stars mark the fiducial model, flat and curved.}
\label{fig:charge}
\end{figure*}

Repeating the calculation over a family of polytropes, $1.8\le\Gamma\le3$ with $K$ chosen to span realistic masses and radii, and retaining only models on the stable branch below the maximum mass, gives $150$ stellar models covering $0.065\le\mathcal{C}\le0.30$. Their charges collapse onto a single curve in the compactness,
\begin{equation}
q_\infty \simeq 1 - 7.2\,\mathcal{C}^2 ,
\label{eq:qfit}
\end{equation}
with a root-mean-square residual of $0.03$ and a maximum residual of $0.07$ across the whole family (Fig.~\ref{fig:charge}(b)). The scalar charge is thus quasi-universal, like the I--Love--Q relations~\cite{Yagi:2013awa} and in the sense of Yagi \& Stepniczka~\cite{YagiStepniczka2021}: it is fixed by the compactness to a few per cent, and is otherwise insensitive to the \EoS. This answers, in the affirmative and with a number attached, the question of whether $q$ is \EoS-sensitive at all. It is also an instance of a broader pattern: Yagi \& Stepniczka~\cite{YagiStepniczka2021} found a relation between scalar charge and binding energy, insensitive to the \EoS, in the massless Damour--Esposito-Far\`ese theory; since $q=1+\Omega/M$ in the Newtonian limit, $\Omega$ being the gravitational binding energy, our relation is a binding-energy relation of the same kind, specialised to the scalaron. The novelty here is the massive case, the explicit factor $(1-\mathcal{C})$ of Eq.~(\ref{eq:redshift}), and the finite-range correction of Eq.~(\ref{eq:qlam}).

The finite range of the scalaron enters only at second order in $R_\star/\lam$, since the kernel of Eq.~(\ref{eq:qdef}) expands as $1+(m_\phi r)^2/6+\dots$, and we find empirically
\begin{equation}
q(\mathcal{C},R_\star/\lam)\simeq q_\infty(\mathcal{C}) \left[1+0.076\,(R_\star/\lam)^2\right] ,
\label{eq:qlam}
\end{equation}
accurate to better than two per cent for $R_\star\lesssim\lam$. The correction is not a curiosity: it is positive, so that a sufficiently short scalaron range restores to the star the charge that pressure took from it. For the fiducial model $q=0.78$ at $\lam\gg R_\star$, $0.83$ at $\lam=R_\star$ and $1.01$ at $\lam=R_\star/2$, so that the bound $q\le1$, which assumes $R_\star\ll\lam$, is exceeded for $\lam\lesssim0.52\,R_\star$. Both statements, the pressure deficit and the finite-size enhancement, are visible in Fig.~\ref{fig:charge}(b).

\subsection{The interior propagator on the curved background}
\label{sec:nscurved}

Equation~(\ref{eq:charge}) draws its source from a relativistic star but propagates the field with the flat-space kernel, which is not how the disc of Sec.~\ref{sec:bh} was treated. The remedy is already written down: we evaluate Eq.~(\ref{eq:master}) on the interior metric, that is Eq.~(\ref{eq:curvedradial}) with $m(r)$ the enclosed mass and $\Phi$ fixed by $d\Phi/dr=(m+4\upi r^3p)/[r(r-2m)]$, matched to $\tfrac12\ln(1-2M/R_\star)$ at the surface. The regular mode $u$ is integrated outwards from the centre and the decaying mode $v$ inwards from large radius, and the charge follows from Eq.~(\ref{eq:master}).

The effect is a systematic reduction, and not a small one. For the fiducial star $q=0.66$, against the $0.78$ of Eq.~(\ref{eq:qinf}), and across the polytrope family the two determinations stand in the ratio
\begin{equation}
\frac{q_{\rm curved}}{q_{\rm flat}} \simeq 1-\mathcal{C} ,
\label{eq:redshift}
\end{equation}
to about two per cent over $0.04\le\mathcal{C}\le0.24$, tending to unity in the Newtonian limit. In the long-range limit its origin is exact: there $u=1$ and $|W|=1$ on either background, so the ratio is that of the two source measures, and the volume factor $e^{\Lambda}$ and the redshift $e^{\Phi}$, which cancel at the surface, leave inside the star a residual $\langle\Phi+\Lambda\rangle\simeq-\mathcal{C}$. Combining Eq.~(\ref{eq:redshift}) with Eq.~(\ref{eq:qfit}) gives
\begin{equation}
q_{\rm curved} \simeq (1-\mathcal{C})\left(1-7.2\,\mathcal{C}^2\right) , \qquad \lam\gg R_\star ,
\label{eq:qfitcurved}
\end{equation}
which matches the curved polytrope family to an rms residual of $0.032$ and is linear in the compactness at leading order, where the flat-space relation is quadratic. The Newtonian limit fixes that slope: by the virial theorem $q_{\rm flat}\to1-3\mathcal{C}/(5-n)$ for index $n$, and since the equation of state enters only through $3/(5-n)$, which varies by some twenty per cent across the family, the relation is quasi-universal.

Inside the fiducial star the scalaron reaches $|\phi|_{\max}=0.18$, and across the family $|\phi|_{\max}/\mathcal{C}$ falls from $1.24$ to $1.06$; at long range $\phi\simeq-\tfrac23\Phi_{\rm N}$, two-thirds of the depth of the potential well. The neglected back-reaction is therefore of order $\mathcal{C}$, the order of Eq.~(\ref{eq:redshift}) itself, and Sec.~\ref{sec:scf} treats it explicitly; Sec.~\ref{sec:eos} first tests the relation against realistic equations of state, which changes the coefficient of Eq.~(\ref{eq:qfitcurved}) but not its form.

\subsection{Realistic equations of state and the refit}
\label{sec:eos}

The polytropes are stand-ins, so we repeat the calculation with six equations of state that span the soft-to-stiff range and two compositions: \textsc{apr4}~\cite{Akmal:1998cf}, \textsc{sly4}~\cite{Douchin:2001sv}, \textsc{mpa1}~\cite{Muther:1987xaa}, \textsc{tm1}~\cite{Sugahara:1993wz}, the hyperonic \textsc{h4}~\cite{Glendenning:1991es}, and the unified relativistic-mean-field model \textsc{uci}a of Malik et al.~\cite{Malik:2024nva,Malik:2024zenodo}, calibrated to nuclear and $\chi$EFT constraints. The first five are piecewise polytropes in the parametrisation of Read et al.~\cite{Read:2008iy}, with the fits of Kumar and Landry~\cite{Kumar:2019xgp} for \textsc{sly4}, \textsc{mpa1} and \textsc{tm1}, and reproduce the published maximum masses. For each we evaluate, at fixed compactness,
\begin{equation}
q_{\rm curved}=1-\frac{6}{M}\int_0^{R_\star}p\,e^{\Phi+\Lambda}\,4\upi r^2\,dr=1-\frac{6P_T}{M},
\label{eq:qcurvedPT}
\end{equation}
with $M$ the Tolman mass.

\begin{figure}
\centering
\includegraphics[width=\columnwidth]{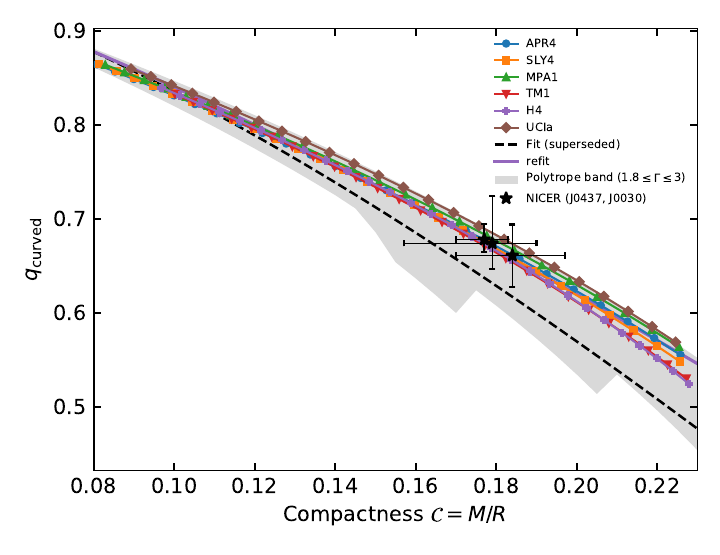}
\caption{Curved-background charge against compactness for the equations of state, the polytrope fit of Eq.~(\ref{eq:qfitcurved}) (dashed), the refit of Eq.~(\ref{eq:refit}) (purple; dotted beyond the fitted range) and the NICER pulsars of Sec.~\ref{sec:data_nicer} (stars, with 68\% intervals). The stars are placed by their measured compactness: the figure shows where measured stars sit on the relation (not an independent test)}
\label{fig:qCmerged}
\end{figure}

The relation survives and the coefficient does not. The spread of $q_{\rm curved}$ at fixed $\mathcal{C}$ grows from $1.2$ per cent at $\mathcal{C}=0.10$ to $5.4$ per cent at $0.22$, \textsc{tm1} and \textsc{h4} falling below the others at the high end (Table~\ref{tab:spread}), so the quasi-universality of Sec.~\ref{sec:ns} holds. The offset from the polytrope fit of Eq.~(\ref{eq:qfitcurved}) is of a different kind: it is one-sided, the realistic curves lying above it at every $\mathcal{C}$ sampled, and the fit falls $10$ per cent below their mean at $\mathcal{C}=0.22$. The reason is the shape of the fit: its slope $(1-q)/\mathcal{C}$ tends to $1$ as $\mathcal{C}\to0$, below the Newtonian value $6/(5-n)$ of every polytrope, whereas the realistic curves give $1.64$ at $\mathcal{C}=0.08$, rising to $1.97$ at $0.22$, so that the polytrope fit bends too early. Against the band of the $1.8\le\Gamma\le3$ polytropes the six-equation mean lies in its upper part for $\mathcal{C}\gtrsim0.11$, with \textsc{tm1} and \textsc{h4} falling to mid-band by $\mathcal{C}=0.22$ (Fig.~\ref{fig:qCmerged}); a coefficient tuned to the whole family therefore undershoots real nuclear matter.

A least-squares fit of the form $(1-\mathcal{C})(1-a\,\mathcal{C}-b\,\mathcal{C}^2)$ to the six-equation mean over $0.10\le\mathcal{C}\le0.225$ gives $(a,b)=(0.205,4.61)$, and we adopt
\begin{equation}
q_{\rm curved}\simeq(1-\mathcal{C})\left(1-0.205\,\mathcal{C}-4.61\,\mathcal{C}^2\right),
\label{eq:refit}
\end{equation}
which follows the mean to an rms residual of $0.004$ and a maximum of $0.007$, individual equations of state lying within $\pm0.024$ of it. Equation~(\ref{eq:refit}) is the relation we use wherever a number is quoted for a real star, in place of the polytrope fit of Eq.~(\ref{eq:qfitcurved}). Like Eq.~(\ref{eq:qfitcurved}), it is a fit over $0.10\le\mathcal{C}\le0.225$ and not an expansion about $\mathcal{C}=0$.

\begin{table}
\centering
\small
\setlength{\tabcolsep}{3pt}
\caption{$q_{\rm curved}$ at fixed compactness for the six equations of state, against the polytrope fit of Eq.~(\ref{eq:qfitcurved}) (``Poly.'') and the refit of Eq.~(\ref{eq:refit}).}
\label{tab:spread}
\begin{tabular}{ccccccccc}
\hline
$\mathcal{C}$ & \textsc{apr4} & \textsc{sly4} & \textsc{mpa1} & \textsc{tm1} & \textsc{h4} & \textsc{uci}a & Poly. & Refit \\
\hline
0.10 & 0.831 & 0.832 & 0.835 & 0.833 & 0.834 & 0.841 & 0.835 & 0.840 \\
0.13 & 0.776 & 0.776 & 0.780 & 0.775 & 0.777 & 0.787 & 0.764 & 0.779 \\
0.16 & 0.715 & 0.714 & 0.720 & 0.711 & 0.714 & 0.727 & 0.685 & 0.713 \\
0.19 & 0.647 & 0.644 & 0.654 & 0.638 & 0.640 & 0.659 & 0.599 & 0.644 \\
0.22 & 0.571 & 0.564 & 0.578 & 0.553 & 0.551 & 0.582 & 0.508 & 0.571 \\
\hline
\end{tabular}
\end{table}

\subsection{The trace anomaly}
\label{sec:anomaly}

\begin{figure}
\centering
\includegraphics[width=\columnwidth]{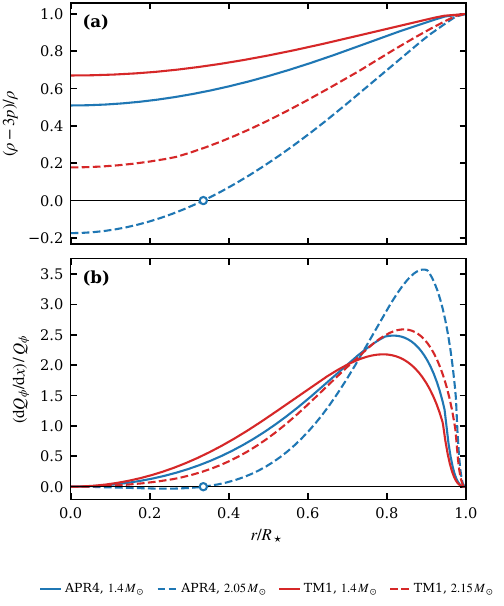}
\caption{Left: the normalised trace anomaly $(\rho-3p)/\rho$ against $r/R_\star$ for \textsc{apr4} (blue) and \textsc{tm1} (red) at $1.4\,\Msun$ (solid) and at $2.05$ and $2.15\,\Msun$ (dashed). Right: the integrand of $Q_\phi$ (Eq.~\ref{eq:qcurvedPT}), normalised so that it integrates to unity over $x=r/R_\star$; the negative core of the massive \textsc{apr4} star is visible and that of \textsc{tm1} is absent.}
\label{fig:anomaly}
\end{figure}

The source of the scalaron is $\rho-3p$, the trace anomaly of the matter, and it need not be positive. In a conformal fluid it vanishes, for a stiff equation of state the pressure can exceed $\rho/3$, and $q_{\rm curved}$ then measures an \emph{integrated} trace anomaly, $3Q_\phi=M-6P_T$ in the notation of Eq.~(\ref{eq:qcurvedPT}). This ties the charge to a live question in dense-matter physics, the behaviour of the trace anomaly in the cores of massive stars~\cite{Fujimoto:2022ohj,Bedaque:2014sqa}. Figure~\ref{fig:anomaly} shows $(\rho-3p)/\rho$ and the integrand of $Q_\phi$ for \textsc{apr4} and \textsc{tm1} at $1.4\,\Msun$ and at a mass close to their maxima.

The result is not the one that stiffness alone would suggest. The source changes sign only for equations of state that are soft at nuclear density and stiffen above it, as current constraints from massive stars suggest~\cite{Annala:2019puf}: $p_c/\rho_c$ crosses $1/3$ near $2\,\Msun$ for \textsc{apr4}, \textsc{sly4}, \textsc{mpa1} and \textsc{uci}a, whereas \textsc{tm1} and \textsc{h4}, stiff throughout, never reach it, and every $1.4\,\Msun$ star has a positive source. Where the models are trustworthy the effect on the charge is small: the negative core contributes $-0.6$ per cent of $Q_\phi$ for \textsc{apr4} at $2.05\,\Msun$, and $-0.7$ per cent for the heaviest stable \textsc{uci}a star, the one causal tabulated model. Within $0.5$ per cent of the maximum mass the piecewise fits give $-5$ to $-21$ per cent, but those stars are superluminal at the centre, so the extreme values reflect the parametrisation as much as nuclear matter.

\subsection{Relations to the tidal deformability and the moment of inertia}
\label{sec:universal}

\begin{figure}
\centering
\includegraphics[width=\columnwidth]{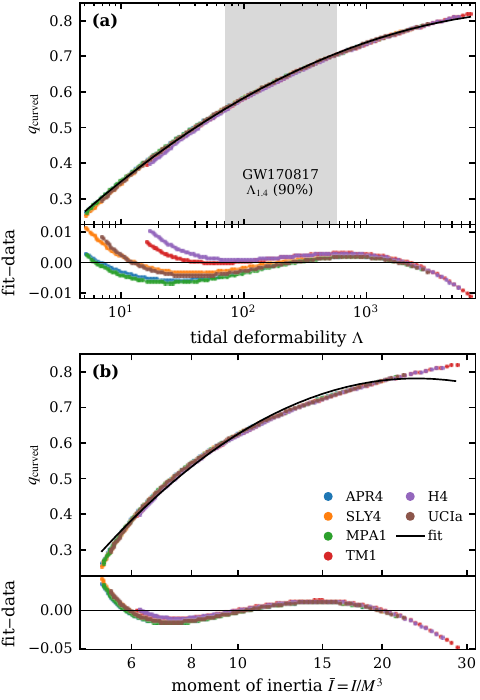}
\caption{The charge against the tidal deformability (left) and the moment of inertia (right) for stable stars of the six equations of state with $M\ge1.0\,\Msun$, with the quadratic fits of the text (black) and residuals below. The grey band on the left is the GW170817 interval for $\Lambda_{1.4}$.}
\label{fig:universal}
\end{figure}

Because $q_{\rm curved}$ is itself an equation-of-state-insensitive function of the compactness, it should correlate with the other quasi-universal stellar quantities, and we compute it against the dimensionless tidal deformability $\Lambda$ (with $\ell=2$ Love number integrated alongside the structure equations~\cite{Hinderer:2007mb,Damour:2009vw,Postnikov:2010yn}) and the dimensionless moment of inertia $\bar I=I/M^3$ (slow rotation~\cite{Hartle:1967ApJ}), over the stable branch for $M\ge1.0\,\Msun$ of the six equations of state. A quadratic in the logarithm,
\begin{equation}
\begin{aligned}
q_{\rm curved}&\simeq c_0+c_1\ln\Lambda+c_2(\ln\Lambda)^2,\\
(c_0,c_1,c_2)&=(0.0372,\,0.1523,\,-0.00734),
\end{aligned}
\label{eq:qLam}
\end{equation}
reproduces the charge to an rms residual of $0.0033$ and a maximum of $0.011$ over $578$ stars, and to $0.0025$ for $M\le1.6\,\Msun$; the analogous fit in $\bar I$ has rms $0.012$ and maximum $0.047$ (Fig.~\ref{fig:universal}).

The point of the relation is that, unlike the NICER mapping of Sec.~\ref{sec:data_nicer}, it does not pass through the compactness that the charge is itself a function of: $\Lambda$ is measured from the waveform. For GW170817, $\Lambda_{1.4}=190^{+390}_{-120}$ (90 per cent~\cite{LIGOScientific:2018cki}) maps through Eq.~(\ref{eq:qLam}) to $q_{\rm curved}=0.63$ with a range $0.55$--$0.71$. This is a second, independent route to the stellar charge of a $1.4\,\Msun$ star, in agreement with the NICER values of Sec.~\ref{sec:data_nicer}, and like them a statement about what the theory predicts for the charge and not a test of the theory itself, which Sec.~\ref{sec:data_gw} brings to the dipole bound.

\subsection{Back-reaction of the scalaron on the star}
\label{sec:scf}

Everything above treats the scalaron as a test field, and with $|\phi|\sim\mathcal{C}$ that cannot be taken on trust. We therefore solve the Einstein-frame structure equations of Damour and Esposito-Far\`ese~\cite{DamourEspositoFarese1993} with the $R+\alpha R^2$ potential and coupling $A=e^{-\varphi/\sqrt3}$, shooting on the central value of the scalar, for \textsc{apr4}, \textsc{sly4} and \textsc{tm1} at the central pressure of the $1.4\,\Msun$ star and $\lam=30$, $100$ and $300$~km (Table~\ref{tab:scf}). The scalaron lowers the mass by up to $3$ per cent for \textsc{apr4} and \textsc{sly4} and $10$ per cent for \textsc{tm1}, and the compactness by $2$--$8$ per cent. Since $q$ is a function of the compactness, the relevant comparison is at fixed $\mathcal{C}$: the star moves along Eq.~(\ref{eq:refit}), and its departure from the test-field relation, $-0.7$ to $-3.8$ per cent, is comparable to the spread between equations of state. The form of the relation therefore survives the back-reaction. The calculation is coarse, one mass and three equations of state, and does not test the maximum-mass region.

\begin{table}
\centering
\small
\setlength{\tabcolsep}{4pt}
\caption{Back-reaction of the scalaron on a star of central pressure equal to that of the $1.4\,\Msun$ general-relativistic star. $\Delta q/q$: charge relative to the test-field value at the same central pressure; $\Delta\mathcal{C}/\mathcal{C}$: Jordan-frame compactness; $\delta_{\rm rel}$: the charge relative to the test-field relation at the star's actual compactness. Coarse: one mass, three equations of state, charge read at $r=3R_\star$.}
\label{tab:scf}
\begin{tabular}{lrrrr}
\hline
EoS & $\lam$ (km) & $\Delta q/q$ (\%) & $\Delta\mathcal{C}/\mathcal{C}$ (\%) & $\delta_{\rm rel}$ (\%) \\
\hline
\textsc{apr4} & 30  & $+1.5$ & $-3.2$ & $-0.7$ \\
              & 100 & $-0.5$ & $-2.7$ & $-2.3$ \\
              & 300 & $-1.0$ & $-2.2$ & $-2.4$ \\
\textsc{sly4} & 30  & $+1.6$ & $-4.4$ & $-1.2$ \\
              & 100 & $-0.5$ & $-3.3$ & $-2.6$ \\
              & 300 & $-0.9$ & $-3.3$ & $-3.0$ \\
\textsc{tm1}  & 30  & $+2.2$ & $-8.2$ & $-1.4$ \\
              & 100 & $-0.2$ & $-7.5$ & $-3.4$ \\
              & 300 & $-0.7$ & $-7.5$ & $-3.8$ \\
\hline
\end{tabular}
\end{table}

\section{Confronting the charge with data}
\label{sec:data}

Two bodies of data bear on the results above: neutron-star radii and tidal deformabilities, which fix the compactness or the stiffness and hence, through Eqs.~(\ref{eq:refit}) and~(\ref{eq:qLam}), the charge; and gravitational-wave observations of neutron-star--black-hole mergers, which bound a scalar dipole. Horizon-scale imaging, which constrains black-hole metrics, is taken up with the black hole in Sec.~\ref{sec:data_eht}. The comparison is a consistency check and not a fit, and what each data set bounds is not in every case the charge $q$ itself; the section says in each case what it bounds instead.

\subsection{Neutron-star radii}
\label{sec:data_nicer}

NICER measures compactness, and Eq.~(\ref{eq:refit}) turns a compactness into a charge, so recent radius measurements can be read directly in terms of $q$. We take the analyses of three pulsars: PSR~J0437$-$4715~\citep{Choudhury:2024xbk} with an independent analysis by Miller et al~\cite{Miller:2025qfq}; PSR~J0740+6620~\citep{Salmi:2024aum} with an independent analysis by Dittmann et al~\cite{dittmann2024precisemeasurementradiuspsr}; and PSR~J0030+0451~\citep{Vinciguerra_2024}, whose data admit two solutions and which was first analysed by Riley et al~\cite{Riley2019} and Miller et al~\cite{Miller_2019}. For J0030+0451 we take the compactness directly from the two solutions of Vinciguerra et al., as quoted by Luo et al.~\cite{luo2024bulkpropertiespsrj00300451}, since mass and radius are strongly correlated in that fit. Table~\ref{tab:nicer} gives the results, with $q$ read from the refit for $\mathcal{C}\le0.225$ and directly from the equations of state for J0740+6620, which lies beyond the fitted range.

\begin{table*}
\centering
\small
\setlength{\tabcolsep}{3pt}
\begin{tabular}{llcccc}
\hline
Pulsar & Analysis & $M$ ($\Msun$) & $R$ (km) & $\mathcal{C}$ & $q_{\rm curved}$ \\
\hline
J0437$-$4715 & Choudhury+24 & $1.418\pm0.037$ & $11.36^{+0.95}_{-0.63}$ & $0.184^{+0.013}_{-0.014}$ & $0.66^{+0.03}_{-0.03}$ \\
             & Miller+26    & $1.418\pm0.044$ & $13.7\pm1.8$           & $0.153^{+0.024}_{-0.018}$ & $0.73^{+0.04}_{-0.05}$ \\
J0740+6620   & Salmi+24     & $2.073\pm0.069$ & $12.49^{+1.28}_{-0.88}$ & $0.244^{+0.021}_{-0.023}$ & $0.49^{+0.07}_{-0.05}$ \\
             & Dittmann+24  & $2.08\pm0.07$   & $12.92^{+2.09}_{-1.13}$ & $0.236^{+0.026}_{-0.032}$ & $0.52^{+0.09}_{-0.08}$ \\
J0030+0451   & ST+PDT       & $1.40^{+0.13}_{-0.12}$ & $11.71^{+0.88}_{-0.83}$ & $0.177^{+0.006}_{-0.007}$ & $0.67^{+0.02}_{-0.01}$ \\
             & PDT-U        & $1.70^{+0.18}_{-0.19}$ & $14.44^{+0.88}_{-1.05}$ & $0.179^{+0.011}_{-0.022}$ & $0.67^{+0.05}_{-0.03}$ \\
\hline
\end{tabular}
\caption{NICER determinations with the compactness and the curved-background charge they imply (medians and 68\% intervals). For J0030+0451, $\mathcal{C}$ is the directly inferred compactness of Vinciguerra et al as quoted by Luo et al, and $M$ and $R$ are listed for reference only. The upper interval for J0740+6620 is truncated at $\mathcal{C}=0.258$, the largest compactness reached by all equations of state on the stable branch; the $q$ values for this pulsar are read directly from the six equations of state. The radius of J0437$-$4715 from Miller et al is the 68\% range $11.9$--$15.5$~km, written as a symmetric interval.}
\label{tab:nicer}
\end{table*}

The question the realistic equations of state can answer, and the polytropes could not, is which limits the comparison, the equation of state or the measurement. For each star we read $q_{\rm curved}$ at the measured $\mathcal{C}$ from the six equations of state and compare their spread with the uncertainty that the NICER posterior propagates into the prediction. Since $q$ is read off at the measured $\mathcal{C}$, this tests the usefulness of the relation, whether its residual equation-of-state dependence is below what can be measured, and not the theory. Table~\ref{tab:nicerspread} uses one standard deviation of the NICER-propagated $q$ against the full equation-of-state range, the conservative convention. For J0437$-$4715 and J0740+6620 the NICER uncertainty exceeds the equation-of-state spread by factors of $1.7$ and $1.6$, and for the PDT-U solution of J0030+0451 by $2.0$; for the ST+PDT solution, whose compactness is constrained to $\pm0.0065$, the factor is $0.8$ and the equation-of-state spread is already the larger. The comparison is therefore measurement-limited in three of the four cases and at parity in the fourth, which marks the point at which the equation-of-state systematic becomes the floor.

\begin{table*}
\centering
\small
\setlength{\tabcolsep}{3pt}
\begin{tabular}{lccccc}
\hline
Pulsar & $\mathcal{C}$ & $q_{\rm curved}$ & EoS range & $\sigma_q$(NICER) & $\sigma_q$/range \\
\hline
J0437$-$4715 & 0.184 & 0.661 & 0.020 & 0.033 & $1.7\times$ \\
J0740+6620 & 0.245 & 0.488 & 0.048 & 0.077 & $1.6\times$ \\
J0030+0451 (ST+PDT) & 0.177 & $0.678^{+0.017}_{-0.013}$ & 0.018 & 0.015 & $0.8\times$ \\
J0030+0451 (PDT-U) & 0.179 & $0.674^{+0.050}_{-0.027}$ & 0.019 & 0.038 & $2.0\times$ \\
\hline
\end{tabular}
\caption{Charge predicted for each pulsar at its measured compactness: mean over the six equations of state, their range (maximum minus minimum), and the NICER-propagated one-standard-deviation uncertainty. For J0437$-$4715 and J0740+6620, $\sigma_q=|dq/d\mathcal{C}|\sigma_{\mathcal{C}}$ with $M$ and $R$ independent and symmetrised radius errors ($\pm0.79$ and $\pm1.08$~km); for J0030+0451, $\mathcal{C}$ is sampled from a split-normal distribution built from the 68.3\% intervals.}
\label{tab:nicerspread}
\end{table*}

\subsection{Gravitational-wave sensitivity to the dipole}
\label{sec:data_gw}

A neutron-star--black-hole binary radiates scalar dipole radiation at $-1$PN order if its members carry unequal charges, a signature also tested with binary pulsars~\cite{Freire:2012mg,Kramer:2021jcw}, and four such events have been used to bound it: GW190814~\citep{LIGOScientific:2020zkf}, GW200105 and GW200115~\citep{LIGOScientific:2021qlt}, and GW230529~\citep{LIGOScientific:2024elc}. Santos et al~\cite{Santos:2025ifr} bound the combination $\kappa_0\hat\alpha^2$ in luminal Horndeski gravity, the tightest bound coming from GW230529, $\kappa_0\hat\alpha^2=(1.0^{+2.5}_{-2.0})\times10^{-4}$ at $90$ per cent confidence, that is $\kappa_0\hat\alpha^2\lesssim3.5\times10^{-4}$ at its upper edge. In Brans--Dicke language, with $s$ the Eardley sensitivity~\citep{Eardley1975,Will:2014kxa} and $\omega_{\rm BD}=0$ for Eq.~(\ref{eq:field}), $\kappa_0\hat\alpha^2=(s_{\BH}-s_{\NS})^2/3$. The black hole has $s_{\BH}=1/2$, its charge vanishing by the no-hair argument of Sec.~\ref{sec:intro}, so that $\kappa_0\hat\alpha^2=(1-2s_{\NS})^2/12$.

The charge of Sec.~\ref{sec:nscurved} is exactly $1-2s_{\NS}$, which is the normalisation of Seymour \& Yagi~\cite{Seymour:2019tir} for quadratic $f(R)$, in which $q=1-2s$ and the coupling is $B=1/3$; what is specific to the present work is its evaluation on the relativistic interior. Writing the Tolman mass as $M=4\upi\int(\rho+3p)\,e^{\Phi+\Lambda}r^2dr$, the long-range limit of Eq.~(\ref{eq:master}) gives
\begin{equation}
q_{\rm curved}=1-\frac{6P_T}{M}=1-2s_{\NS} , \,\, P_T\equiv4\upi\int_0^{R_\star}p\,e^{\Phi+\Lambda}r^2dr ,
\label{eq:qtolman}
\end{equation}
the trace integral that defines the scalar monopole in Brans--Dicke theory~\citep{DamourEspositoFarese1993}. The dipole parameter is therefore
\begin{equation}
\kappa_0\hat\alpha^2=\frac{q_{\rm curved}^2}{12} ,
\label{eq:gwmap}
\end{equation}
which is $0.036$ for $q_{\rm curved}=0.66$, a factor of a hundred above the GW230529 limit. That comparison holds, however, only for a scalar that is massless on the orbital scale. For a massive scalar the dipole flux is multiplied by $(\omega^2-m_\phi^2)/\omega^2$ for $\omega>m_\phi$ and vanishes below it~\citep{Alsing:2011er}, so that there is no dipole emission unless $\lam\gtrsim c/(\upi f)$, with $f$ the gravitational-wave frequency: $4.8\times10^3$~km at $20$~Hz, the lowest frequency of the analysis, and $95$~km at $1$~kHz. The massless analysis of Santos et al~\cite{Santos:2025ifr} is therefore strictly valid only beyond $4.8\times10^3$~km. Below that we use a Fisher-matrix calculation~\cite{Cutler:1994ys,Vallisneri:2007ev} for a GW230529-like system ($3.6+1.4\,\Msun$, signal-to-noise ratio $11.6$ in an aLIGO-like detector~\cite{LIGOScientific:2014pky}, TaylorF2 phase to second post-Newtonian order~\cite{Buonanno:2009zt}, free chirp mass, mass ratio and coalescence time and phase), in which the $-1$PN dipole term~\cite{Yunes:2009ke} carries the cut-off factor $g(f)=1-(f_c/f)^2$ for $f>f_c=c/\upi\lam$ and vanishes below it. The published massless bound, $\kappa_0\hat\alpha^2<3.5\times10^{-4}$, is assigned to the massless limit, and the bound at finite $\lam$ is that value scaled by the ratio of the Fisher uncertainties on the dipole amplitude, $q_{\max}(\lam)=0.065\,[\sigma_B(\lam)/\sigma_B(0)]^{1/2}$. It is $0.86$ at $\lam=300$~km, $0.16$ at $10^3$~km and $0.076$ at $10^4$~km, and is close to $0.065$ beyond a few thousand kilometres, with structure of order a factor of two where $f_c$ crosses the $20$~Hz edge of the band, which a full Bayesian analysis would smooth. On this calibrated estimate, a similar binary is sensitive to a charge $q=0.66$ for $\lam\gtrsim3.5\times10^2$~km, and to $q=0.49$ for $\lam\gtrsim4.0\times10^2$~km.

For third-generation detectors we scale the noise coarsely, with an amplitude spectral density ten times below the aLIGO curve above $10$~Hz, a power-law wall below it, and lower cut-offs of $5$~Hz (Cosmic Explorer~\cite{Evans:2021gyd}) and $3$~Hz (Einstein Telescope~\cite{Hild:2010id}). The same source is then measured at a signal-to-noise ratio near $120$, the massless bound on $q$ falls to $0.007$ and $0.005$, and the sensitivity threshold for $q\simeq0.5$--$0.7$ moves to $\lam\simeq220$--$250$~km. That reaches into the range of $30$--$300$~km used for the black-hole figures of this article, and is the first projection in which the dipole channel could constrain it. The noise curves are scalings and not the published ones, and the projection is a statement about sensitivity and not a forecast of an event rate. The dipole channel is thus sensitive to the scalaron range, down to a few hundred kilometres, and not to the charge. 

A scalaron massless over the Solar system, the case to which the analysis strictly applies, is in any case excluded by the Cassini bound on the post-Newtonian parameter $\gamma$, which it would set to $1/2$~\citep{Bertotti2003tgr,Sotiriou2010}.

\begin{figure}
\centering
\includegraphics[width=\columnwidth]{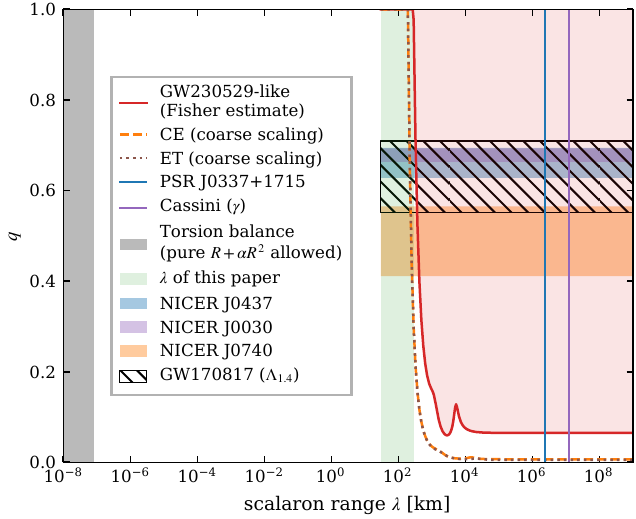}
\caption{Constraints in the $(\lam,q)$ plane. Grey (left): pure $R+\alpha R^2$ is allowed by the torsion-balance bound only for $\lam\lesssim10^{-4}$~m. Red: the Fisher estimate $q_{\max}(\lam)$ of the sensitivity of a GW230529-like binary, calibrated on the published massless dipole constraint and including the massive-dipole cut-off (shaded above); orange and brown dashed: coarse Cosmic Explorer and Einstein Telescope scalings of the same calculation. Vertical lines: PSR~J0337+1715 ($\lam<2.4\times10^6$~km, Ref.~\citep{Seymour:2019tir}, Sec.~\ref{sec:data_eht}) and Cassini ($\lam\lesssim1.2\times10^7$~km, an order-of-magnitude estimate: the delay of light is blind to the Yukawa term, which cancels in $\Phi+\Psi$, whereas the mass inferred from planetary orbits includes it, so the test sees $|\gamma-1|\simeq\tfrac23e^{-x}(1+x)<2.3\times10^{-5}$ with $x=r/\lam$ at $r=1$~AU). Coloured bands: NICER charges of Sec.~\ref{sec:data_nicer}; hatched black band: GW170817 via Eq.~(\ref{eq:qLam}). Green: the $30$--$300$~km range used for the black-hole figures.}
\label{fig:constraints}
\end{figure}

The stellar charge of Sec.~\ref{sec:data_nicer} and Eq.~(\ref{eq:qLam}) feeds directly into the dipole parameter of Eq.~(\ref{eq:gwmap}). Figure~\ref{fig:constraints} collects what bears on the unscreened scalaron in the $(\lam,q)$ plane: the laboratory bound on pure $R+\alpha R^2$ of Sec.~\ref{sec:scope}; the GW230529-calibrated Fisher estimate $q_{\max}(\lam)$ above, with its Einstein Telescope and Cosmic Explorer scalings; the pulsar-timing bound on the range from PSR~J0337+1715~\citep{Seymour:2019tir} and the Cassini bound; the NICER bands of Sec.~\ref{sec:data_nicer}; and the GW170817 band of Sec.~\ref{sec:universal}. The astrophysical constraints and the dipole sensitivity act on $\lam$ and not on $q$: every stellar charge in the figure lies above the dipole limit $q=0.065$, so the dipole channel probes the range and leaves the charge where Eqs.~(\ref{eq:refit}) and~(\ref{eq:qLam}) put it.

\section{The accreting black hole}
\label{sec:bh}

\subsection{Curved against flat background}
\label{sec:curvedflat}

\begin{figure}
\centering
\includegraphics[width=\columnwidth]{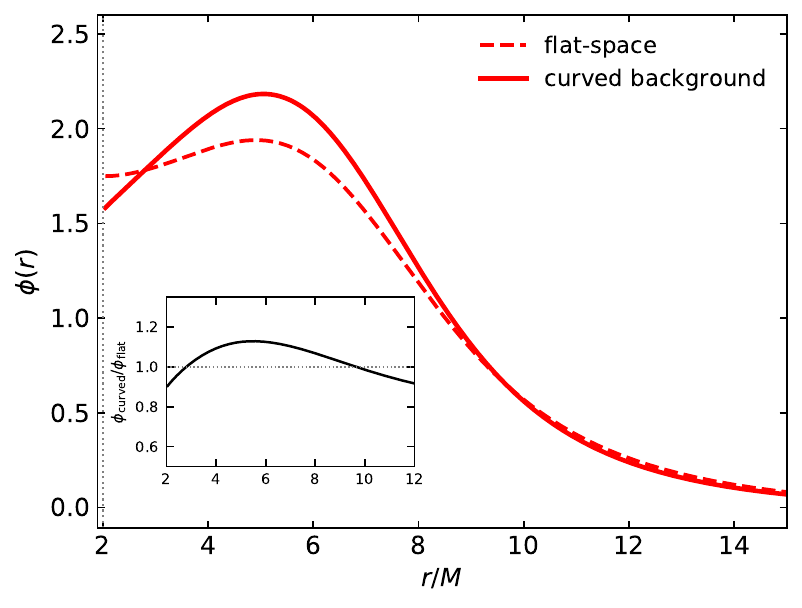}
\caption{Curved-background, horizon-regular scalaron profile against the flat-space stand-in for an accreting black hole, for the toy Gaussian disc at $r_d=6M$ with $m_\phi=0.3/M$. Inset: the ratio $\phi_{\rm curved}/\phi_{\rm flat}$.}
\label{fig:main}
\end{figure}

Figure~\ref{fig:main} compares the curved-background solution with the flat-space stand-in for a toy Gaussian disc centred at $r_d=6M$,
\begin{equation}
\rho(r) = \rho_0\exp\!\left[-\frac{(r-r_d)^2}{2\sigma_d^2}\right] ,
\label{eq:gauss}
\end{equation}
a Gaussian ring being a convenient idealisation of the pressure-supported torus that occupies the inner region of a thick disc~\citep{AbramowiczFragile2013}, and serving here only to localise the source. Here $\rho_0=0.05$ and $\sigma_d=1.5M$, and the outer condition is $\phi(r_{\rm out})=0$ at $r_{\rm out}=20M$ in both cases. The correction is not a constant rescaling: the ratio $\phi_{\rm curved}/\phi_{\rm flat}$ peaks near $1.13$ at the disc, crosses unity near $r\approx9.5M$, and the two solutions decay at the same Yukawa rate in the far field, which is itself the test described below. Over $m_\phi\in[0.15,1.0]/M$, that is $\lam\in[1,6.7]M$, the correction at the disc falls monotonically from about $18$ per cent, for ranges approaching the horizon scale, to about $6$ per cent, so it is generic across the range.

\subsection{A physically motivated disc}
\label{sec:disk}

The toy Gaussian is replaced by a simplified Shakura--Sunyaev/Novikov--Thorne law~\citep{ShakuraSunyaev1973,NovikovThorne1973},
\begin{equation}
\Sigma(r) = \Sigma_0\, r^{-3/4}\left[1-\sqrt{r_\ISCO/r}\,\right] ,
\qquad r\ge r_\ISCO=6M ,
\label{eq:ss}
\end{equation}
converted to a volume density $\rho=\Sigma/2H$ through a thin-disc scale height $H(r)=h_0r$ with $h_0=0.05$, so that $\rho=(\Sigma_0/2h_0)\,r^{-7/4}[1-\sqrt{r_\ISCO/r}\,]$. The amplitude $\Sigma_0$ takes no part in what follows, Eq.~(\ref{eq:qdef}) being normalised by the source mass, so that $q$ is fixed by the shape of the profile and not by its scale. Unlike the Gaussian, this profile has a sharp inner edge at the innermost stable circular orbit and a slowly decaying tail instead of a localised bump, its maximum lying at $r=(81/49)\,r_\ISCO\approx9.9M$.

The two profiles differ in a way that matters more for the exterior field than the profiles themselves suggest. Evaluating Eq.~(\ref{eq:qdef}) for the Gaussian disc gives $q=1.90$, a well-converged number because the source is localised. For the Shakura--Sunyaev profile, whose tail decays only as a power law, the $\sinh$ kernel wins and the charge is set by wherever the disc is truncated: $q=13.1$ for an outer edge at $20M$, $35.4$ at $25M$ and $104$ at $30M$, for $m_\phi=0.3/M$. The exterior scalar charge of a realistic disc is therefore controlled by its outer regions and any quantitative prediction must state the truncation radius alongside the accretion rate. We regard this as one of the more useful outcomes of replacing the toy profile, and it is visible in Fig.~\ref{fig:charge}(a) as the departure of the Shakura--Sunyaev points from the thin-shell curve.

\subsection{Validity of the disc model}
\label{sec:validity}

The disc results rest on two approximations, quasi-stationarity and the reduction of the disc to a shell; the first is safe, and the second is not small.

\paragraph*{Quasi-stationarity.}
\label{sec:quasistatic}

For standard alpha-disc parameters~\citep{ShakuraSunyaev1973,FrankKingRaine2002}, $\alpha_{\rm visc}=0.1$ and $H/r=0.05$ at $r_d=6M$, the viscous timescale is $t_{\rm visc}=t_{\rm dyn}/[\alpha_{\rm visc}(H/r)^2]\approx 1.1\times10^5\,t_\phi$, where $t_{\rm dyn}=2\upi(r_d^3/M)^{1/2}$ is the orbital period at the disc and $t_\phi=1/m_\phi$ is the scalaron relaxation time.  Across the entire range explored, including a geometrically thick, advection-dominated state with $H/r\sim1$~\citep{NarayanYi1994,YuanNarayan2014}, the ratio never falls below about $7$; the quasi-stationary assumption is therefore safe throughout and not merely at the fiducial point, and it becomes marginal only where high viscosity and extreme thickness occur together.

\paragraph*{Reduction of the disc to a shell.}
\label{sec:monopole}

The disc is reduced throughout to a spherically symmetric shell at $r_d$. That reduction is exact for the monopole: the $\ell=0$ component of any source is its angular average, and the angular average of matter at $r_d$ is a shell at $r_d$. Since the kernel of Eq.~(\ref{eq:qdef}) depends on radius alone, matter contributes the same charge wherever it lies in angle, so $q$ is untouched and what the reduction discards is entirely the $\ell\ge1$ content. The rest of this section measures that.  The numbers quoted are for $r_d=6M$ and $m_\phi=0.3/M$, the fiducial of Sec.~\ref{sec:curvedflat}. The angular weight of that truncation is set by $P_\ell(\cos\theta=0)$ for a source confined to the equatorial plane, $P_\ell$ being the Legendre polynomials. The odd multipoles vanish identically there, and the even ones give $|P_\ell(0)/P_0(0)|=1$, $1/2$, $3/8$, $5/16$ and $35/128$ for $\ell=0$ to $8$, falling only as $(2/\upi\ell)^{1/2}$.

The Yukawa multipole expansion of the Green's function~\citep{Olver2010} gives, for a ring source at $r_d$, the exterior quadrupole-to-monopole ratio in the plane of the disc, the worst case and the one relevant to a grazing photon,
\begin{equation}
\eta(r) = \frac{5}{4}\, \frac{i_2(m_\phi r_d)}{i_0(m_\phi r_d)}\, \frac{k_2(m_\phi r)}{k_0(m_\phi r)} ,
\label{eq:Q}
\end{equation}
with $i_\ell$ and $k_\ell$ the modified spherical Bessel functions. Because $k_\ell(x)/k_0(x)\to1$ as $x\to\infty$ for every $\ell$, all multipoles of a massive field share the same exponential decay, unlike the massless case: $\eta$ falls from $0.74$ at the disc only to $0.26$ by $r\sim40M$ and tends to $0.21$, never approaching zero. \emph{The monopole reduction is therefore an $O(1)$ simplification at every radius probed here}; a complete treatment would require the $\ell=2$ correction to both the field equation and the deflection formula.

\subsection{Lensing and the ordering of the charges}
\label{sec:optics}

In the weak field the scalaron enters the two metric potentials with opposite signs~\citep{Sbisa2019,KumaranLopes2026}, $\Phi=-M/r-\phi/2$ and $\Psi=-M/r+\phi/2$, so that the combination $\Phi+\Psi=-2M/r$ that governs null geodesics is exactly the general-relativistic one, whatever the source of $\phi$. The leading deflection of an accreting black hole is therefore $4M/b$, with $b$ the impact parameter, as for a star or a vacuum hole. The scalaron survives only at second order, through the \GW\ \GB\ integral~\citep{Gibbons2008,KumaranLopes2026PRD}, as a fractional deviation
\begin{equation}
\delta\hat\alpha_{\rm frac}(b) = \frac{\sqrt{2\upi}}{3}\;q\; \frac{M_{\rm src}}{\lam}\sqrt{\frac{\lam}{b}}\;e^{-b/\lam} ,\qquad b\gg\lam ,
\label{eq:fracdev}
\end{equation}
which reduces to the point-source result of Ref.~\cite{KumaranLopes2026} for $q=1$ and $M_{\rm src}=M$; everything new is carried by $q$ (Table~\ref{tab:charges}).

 A vacuum black hole has $q=0$ and no correction at any order. A $1.4\,\Msun$ neutron star has $q=0.78$, or $0.66$ for the fiducial polytrope with the curved interior propagator of Sec.~\ref{sec:nscurved} ($0.66$--$0.74$ across the six realistic equations of state of Sec.~\ref{sec:eos}), and lies below the point-mass benchmark either way. An accreting black hole has $q=\sinh(m_\phi r_d)/(m_\phi r_d)$, which exceeds unity and grows exponentially with the radius of the disc, so that a hole with a disc at $r_d=3\lam$ deflects light $4.3$ times more strongly, in the screened correction, than a neutron star of the same mass, or $5.1$ times if the curved-propagator charge is used for the star. The ordering of the three cases is not the one intuition suggests: the object with no hair of its own can carry the largest scalar signature, provided it is not alone.

\begin{table}
\centering
\small
\setlength{\tabcolsep}{4pt}
\begin{tabular}{lcc}
\hline
Environment & $q=3Q_\phi/M_{\rm src}$ & Set by \\
\hline
Vacuum black hole & $0$ & no-hair theorem \\
Neutron star, $1.4\,\Msun$ & $0.78$* & pressure, Eq.(\ref{eq:qinf}) \\
Point mass (referent) & $1$ & definition \\
Gaussian disc, $r_d=6M$ & $1.90$ & $\sinh$ weighting \\
Shakura--Sunyaev disc & $13-10^2$ & outer truncation \\
\hline
\end{tabular}
\caption{The dimensionless scalar charge of Eq.~(\ref{eq:qdef}) in our environments considered, for $m_\phi=0.3/M$ in the disc cases and $\lam\gg R_\star$ for the star; *($0.66-0.74$): the value is the stellar charge with the curved interior propagator of Sec.~\ref{sec:nscurved}. The ordering, and not merely the magnitude, is the result: matter far from the centre is weighted exponentially, pressure is weighted negatively, and vacuum carries nothing.}
\label{tab:charges}
\end{table}

\subsection{Astrophysical amplitude}
\label{sec:amplitude}

The amplitude calls for caution. The fiducial Gaussian disc has $M_{\rm disc}\approx90\,M$, a self-gravitating torus rather than an accretion disc, for which the test-field treatment on a fixed Schwarzschild background would be inconsistent. Because Eq.~(\ref{eq:field}) is linear, every profile shape, every ratio and the charge $q$ itself are independent of $\rho_0$; only the absolute amplitude of $\phi$ scales with it. With the toy normalisation the field reaches $|\phi|\sim2$ at the disc, well outside the regime $|\phi|\ll1$ in which the fixed background is legitimate. Rescaled by the twenty or so orders of magnitude needed to reach a genuine accretion regime, the disc of Sec.~\ref{sec:curvedflat} ($q=1.90$, $\lam=10M/3$) gives, from Eq.~(\ref{eq:fracdev}) at $b=10\lam$, $\delta\hat\alpha_{\rm frac}\sim6\times10^{-25}$, and even the toy amplitude gives only $6\times10^{-4}$. Pushed inwards to the photon ring, $b=3\sqrt3\,M$, the toy amplitude would return $\delta\hat\alpha_{\rm frac}\sim7$; that this exceeds unity is the deflection formula agreeing that this amplitude lies outside the test-field regime. No self-consistent disc therefore yields an observable lensing correction. The stellar channel needs no such rescaling, its field reaching only $|\phi|$ of order $\mathcal{C}$ (Sec.~\ref{sec:nscurved}).

\subsection{Horizon-scale imaging}
\label{sec:data_eht}

Sgr~A$^*$ and M87$^*$ are low-accretion-rate systems, in which the scalaron is expected to be unscreened~\citep{Frolov:2017asg}. Published shadow bounds on its mass, such as that of Kalita \& Bhattacharjee~\cite{Kalita:2023xlu}, refer to a metric whose Yukawa term is sourced by the hole's own mass, a hairy black hole with $q=1$. Here $q_{\BH}=0$, the shadow is general-relativistic at any $\lam$, and the disc correction is negligible (Sec.~\ref{sec:amplitude}). The bound that does apply, from PSR~J0337+1715~\cite{Ransom:2014xla,Archibald:2018oxs,Seymour:2019tir}, $\lam\lesssim2.4\times10^{6}$~km, is weaker than the dipole sensitivity estimated in Sec.~\ref{sec:data_gw}. For comparison, the observed image size of Sgr~A$^*$ agrees with the Kerr prediction to within about $10$ per cent~\cite{EHT2022SgrAVI}.

\section{Discussion}
\label{sec:discussion}

The black-hole result belongs to the line of work, recalled in Sec.~\ref{sec:intro}, showing that matter around a black hole sources a scalar without violating the no-hair theorem~\cite{Jacobson1999,HorbatschBurgess2012,Berti2013,Cardoso2013}; what we add is its computation for a specific theory and environment, with a horizon-regular boundary condition, reduced to the single number $q$.

On the stellar side, interiors in $R+\alpha R^2$ gravity have been constructed both perturbatively and self-consistently~\citep{Cooney2010,Yazadjiev2014,AstashenokCapozzielloOdintsov2014,AstashenokOdintsovDombriz2017}, and the pressure weighting of the weak-field scalar charge was obtained by Sbis\`a~\cite{Sbisa2019}. Our contribution here is not the weighting but its evaluation: Eqs.~(\ref{eq:qinf}) to~(\ref{eq:qlam}) turn it into a number and a quasi-universal relation in the compactness, and identify the regime $\lam\lesssim R_\star/2$ in which the finite-size enhancement overturns the pressure deficit. We are not aware of a previous statement of either result.

Three points need stating plainly. First, the scalaron range is a free parameter here, and the laboratory bound on pure $R+\alpha R^2$ lies some nine orders of magnitude below the ranges that matter astrophysically (Sec.~\ref{sec:scope}); the results apply to the unscreened scalaron of the field equation, Eq.~(\ref{eq:field}), and to a completion of it that evades the laboratory bound, and not to pure $R+\alpha R^2$ as a viable theory. Second, the back-reaction of the scalaron on the stellar structure is of the order $|\phi|\sim\mathcal{C}$; the coarse calculation of Sec.~\ref{sec:scf} shows it moves a star along Eq.~(\ref{eq:refit}) and leaves it within $4$ per cent of the test-field value at fixed compactness, but for one mass and three equations of state, and it is not a replacement for a full treatment of the maximum-mass region. Third, the equation-of-state set is not exhaustive, the crust of each piecewise-polytrope equation is the single fixed segment of Read et al.\ which we have not varied, and the heaviest stars of the piecewise fits are acausal at the centre (Sec.~\ref{sec:anomaly}).

\section{Conclusion}
\label{sec:conclusion}

The exterior charge of the Starobinsky scalaron is an integral over the surrounding matter: the trace of the stress--energy weighted by the interior regular mode of the wave operator, which is $\sinh(m_\phi r)/(m_\phi r)$ in flat space and the regular solution of Eq.~(\ref{eq:curvedradial}) on a curved background. A vacuum horizon contributes nothing, which is the no-hair theorem restated, and the same integral fixes the field of a neutron star.

For a neutron star the pressure subtracts, and on the interior metric $q_{\rm curved}=1-6P_T/M=1-2s$. The charge is a quasi-universal function of the compactness, $q_{\rm curved}\simeq(1-\mathcal{C})(1-0.205\,\mathcal{C}-4.61\,\mathcal{C}^2)$ over $0.10\le\mathcal{C}\le0.225$, which lies up to $10$ per cent above the polytrope-calibrated fit of Eq.~(\ref{eq:qfitcurved}) at $\mathcal{C}=0.22$; it is also a function of the tidal deformability to $0.003$. The source $\rho-3p$ changes sign in the cores of massive stars whose equations of state stiffen sharply above nuclear density, with an effect on $q$ that runs from under one per cent at $2\,\Msun$ to tens of per cent in the extreme, partly acausal, fits.

For the data, NICER and GW170817 fix the stellar charge to between $\pm0.03$ and $\pm0.1$, the comparison being measurement-limited in three of four cases; and a Fisher estimate calibrated on the published massless dipole constraint from GW230529, with the massive-dipole cut-off, shows that a similar binary is sensitive to the scalaron range, down to $\lam\simeq3.5\times10^2$~km for $q=0.66$, and not to the charge, since every stellar charge exceeds the dipole limit; third-generation detectors would reach $220$--$250$~km. The laboratory bound on pure $R+\alpha R^2$ is some nine orders of magnitude tighter in $\lam$, so these results hold for the unscreened scalaron of the field equation and not for that theory as it stands. For an accreting black hole, $q=\sinh(m_\phi r_d)/(m_\phi r_d)$ for matter at $r_d$, set by where a disc ends and not where it shines, and light is very nearly blind to all of it. The back-reaction of the scalaron on the star, solved coarsely here, leaves the relation intact to $4$ per cent; extending it to the full family and to the maximum-mass region is the natural next step.

\section*{Acknowledgements}
Y.~K. acknowledges that this work was supported by the Universidad Nacional Aut\'onoma de M\'exico Postdoctoral Program (POSDOC). I.L. thanks the Funda\c{c}\~ao para a Ci\^encia e Tecnologia (FCT), Portugal, for financial support to the Centre for Astrophysics and Gravitation (CENTRA/IST/ULisboa) through grant No.~UID/PRR/00099/2025 (\doi{10.54499/UID/PRR/00099/2025}) and grant No.~UID/00099/2025 (\doi{10.54499/UID/00099/2025}).

\subsection*{\textbf{Data Availability:}}
The published NICER and gravitational-wave results used here are cited in the text. The code that reproduces the figures is available from the corresponding author upon reasonable request.

\bibliographystyle{apsrev4-2}
\bibliography{references}

\end{document}